\documentclass[aps,pra,twocolumn,groupedaddress,showpacs]{revtex4}

\usepackage{graphicx}

\begin{document}

\title{Weak Measurements with Arbitrary Pointer States}

\author{Lars M. Johansen}

\affiliation{Department of Technology, Buskerud University
College, P.O. Box 251, N-3601 Kongsberg, Norway}

\email{lars.m.johansen@hibu.no}

\date{\today}

\begin{abstract}

The exact conditions on valid pointer states for weak measurements
are derived. It is demonstrated that weak measurements can be
performed with any pointer state with vanishing probability
current density. This condition is found both for weak
measurements of noncommuting observables and for $c$-number
observables. In addition, the interaction between pointer and
object must be sufficiently weak. There is no restriction on the
purity of the pointer state. For example, a thermal pointer state
is fully valid.

\end{abstract}

 \maketitle

In ``orthodox" quantum mechanics, the result of a ``measurement"
is always one of the eigenvalues of the observable. In the last
pages of his textbook on quantum mechanics
\cite{Neumann-MathFounQuanMech:55}, von Neumann provided a model
of a measurement where the object under study was interacting with
a measurement pointer. By assuming that the initial uncertainty of
the pointer was small, von Neumann demonstrated that the pointer
would display one of the eigenvalues of the object observable.

Aharonov, Albert and Vaidman (AAV) considered the same
experimental arrangement
\cite{Aharonov+AlbertETAL-ResuMeasCompSpin:88}. However, they made
the opposite assumption, namely that the initial uncertainty of
the pointer was large. They demonstrated that despite this, the
pointer would on average show the correct expectation value of an
observable $\hat{c}$, although it could not distinguish separate
eigenvalues. Their most interesting discovery, though, was that if
a projective measurement of a second observable $\hat{d}$ was made
on the \emph{object} after the interaction, the average meter
reading conditioned on the result of the measurement on the object
would be the real part of the quantity
\begin{equation}
    c_w(d) = {\langle d \mid \hat{c} \mid \psi \rangle
    \over \langle d \mid \psi \rangle}.
\end{equation}
The authors introduced the name ``weak value" for this quantity.
They observed that the values of $c_w$ might lie outside the range
of eigenvalues of the observable $\hat{c}$. It has been contested
whether the experimental arrangement of AAV qualifies as a
``measurement", and whether it has any meaning to ascribe to $c_w$
a significance as a ``value" of the observable $\hat{c}$
\cite{Leggett-CommResuMeasComp:89,%
Peres-QuanMeaswithPost:89,%
Aharonov+Vaidman-AharVaidRepl:89}.

Despite initial scepticism, weak values have found applications in
a variety of systems. A classical optical analog
\cite{Duck+StevensonETAL-senswhicweakmeas:89} of the original
experiment proposed by AAV has been realized experimentally
\cite{Ritchie+StoryETAL-RealMeasWeakValu:91}. The polarization
state of a classical radiation field can be treated as analogous
to a spin-$1 \over 2$ system, and the weak value of polarization
may exceed the eigenvalue range
\cite{Knight+Vaidman-WeakMeasPhotPola:90,%
Parks+CullinETAL-ObsemeasoptiAhar:98}. It has also been found that
weak values have applications within classical optical
communication \cite{Brunner+AcinETAL-Optitelenetwweak:03,%
Solli+McCormickETAL-Fastlighslowligh:03}. These examples
demonstrate that weak values have applications beyond quantum
mechanics. In fact, it is not unfamiliar that the quantum
formalism can be applied even to classical systems where two
observables cannot be jointly measured with arbitrary accuracy.
This is known e.g. in signal processing, where the Wigner
distribution for time and frequency has gained popularity
\cite{Cohen-TimeAnal:95}.

It was recently found that ``weak values" have a deeper
significance when considered from the viewpoint of standard
Bayesian estimation theory \cite{Johansen-Whatvaluobsebetw:03}. If
the preselected state is considered as ``prior information", an
estimate can be made of the observable $\hat{c}$ on basis of the
postselection measurement of the observable $\hat{d}$. It then
turns out that the ``weak value" is the most efficient estimator
for the observable $\hat{c}$. One may, instead of performing a
weak measurement between pre- and postselection, try to guess the
value of $\hat{c}$. The best possible guess between pre- and
postselection is nothing else than the weak value.

Furthermore, it can be shown \cite{Johansen-Noncpropcohestat:03}
that the real part of the weak value can be expressed as a
conditional moment of the Margenau-Hill distribution
\cite{Margenau+Hill-CorrbetwMeasQuan:61}. More generally, the weak
value is a conditional moment of the standard ordered distribution
\cite{Mehta-PhasFormDynaCano:64}, which is the complex conjugate
of the Kirkwood distribution \cite{Kirkwood-QuanStatAlmoClas:33}.
This is a consequence of the noncommutativity of the observables
involved in weak measurements. One may also consider weak
measurements of observables when they are treated as $c$-numbers.
In this case, it can be shown that the resulting weak value is a
conditional moment of a classical joint distribution of the two
observables involved. Weak values are simply the expectation value
of one variable given the second variable.

AAV considered a specific experimental arrangement where the
measurement pointer was initially in a pure gaussian state with a
large uncertainty. One may wonder whether the emerging weak value
is dependent on this specific state of the pointer? It has been
pointed out that weak values can be observed also for some other
pure pointer states \cite{Aharonov+Vaidman-Two-VectFormQuan:02},
although it seems that no full specification of which states are
allowable has been done. It has also been claimed that weak values
are due to an interference effect of the pointer. This seems to
rule out mixed states as viable pointer states in weak
measurements \cite{Aharonov+Vaidman-Two-VectFormQuan:02}.

It is the purpose of this Letter to demonstrate that weak
measurements can be performed with virtually any state of the
pointer. The only necessary restriction is that the current
density of the pointer must vanish. We shall see that this
conclusion is valid both for noncommuting and $c$-number
observables.

We first examine weak measurements of noncommuting observables. We
consider an object and a pointer described by the density
operators $\hat{\rho}_s$ and $\hat{\rho}_a$, respectively. Prior
to the measurement interaction, the combined object plus pointer
is assumed to be in a product state $\hat{\rho}_0 = \hat{\rho}_s
\otimes \hat{\rho}_a$. We wish to perform a weak measurement of an
arbitrary object observable $\hat{c}$. To this end, we shall
assume that the interaction part of the Hamiltonian has the form
\begin{equation}\label{eq:quantuminteraction}
    \hat{H}_\epsilon = \epsilon \delta(t) \; \hat{c} \otimes \hat{P}.
\end{equation}
This interaction Hamiltonian is the one employed by von Neumann
\cite{Neumann-MathFounQuanMech:55}. For a discussion on the
reasons for using exactly this interaction term, see Ref.
\cite{Haake+Walls-Overamplmetequan:87}. $\hat{P}$ is the momentum
observable of the pointer. We will consistently denote observables
associated with the pointer by capital letters. We assume that
during the measurement interaction, the interaction part of the
Hamiltonian dominates the time evolution. Nevertheless, we shall
assume that the interaction between the object and pointer is
weak, i.e., $\epsilon$ is so small that we can perform a series
expansion to first order in $\epsilon$.

The time evolution is determined by (setting $\hbar=1$)
\begin{equation}
    {\partial \rho \over \partial t} = - i [\hat{H}_\epsilon,\rho].
\end{equation}
Because of the interaction between the object and pointer, the
density operator evolves to $\hat{\rho}_\epsilon =
\hat{U}_\epsilon \hat{\rho}_0 \hat{U}_\epsilon^{\dag}$. Since the
Hamiltonian commutes with itself at all times, the evolution
operator $\hat{U}_\epsilon$ can be written as
\begin{equation}
    \hat{U}_\epsilon = e^{- i \int \hat{H}_\epsilon(t) dt } =
    e^{- i \epsilon \hat{c} \otimes \hat{P}}.
\end{equation}
A projective measurement will be made of the pointer position
$\hat{Q}$ and an object observable $\hat{d}$. We therefore
consider the joint probability distribution
\begin{equation}
    \rho_\epsilon(Q,d) = \langle d \mid \otimes \langle Q \mid
    \hat{\rho}_\epsilon \mid Q \rangle \otimes \mid d \rangle.
\end{equation}
We shall study $\rho_\epsilon(Q,d)$ for small $\epsilon$. We
therefore perform a Maclaurin expansion in $\epsilon$. Using the
fact that
\begin{equation}
    \left ( {\partial^n \hat{U}_\epsilon \over \partial
    \epsilon^n} \right )_{\epsilon=0} = (-i)^n \hat{c}^n \otimes
    \hat{P}^n,
\end{equation}
we obtain the expansion
\begin{widetext}
\begin{eqnarray}
    \rho_\epsilon (Q,d) &=& \rho_0(Q,d) +
    i \epsilon \left ( \langle d
    \mid \hat{\rho}_s \hat{c} \mid d \rangle \langle Q \mid \hat{\rho}_a
    \hat{P} \mid Q \rangle - \langle d \mid \hat{c} \hat{\rho}_s \mid
    d \rangle \langle Q \mid \hat{P} \hat{\rho}_a \mid Q \rangle.
    \right ) + R_\epsilon(Q,d)
\end{eqnarray}
where
\begin{equation}
    \rho_0 (Q,d)= \langle d \mid \hat{\rho}_s \mid d \rangle
    \langle Q \mid \hat{\rho}_a \mid Q \rangle
\end{equation}
is the joint probability distribution for $Q$ and $d$ prior to the
interaction, and where
\begin{equation}\label{eq:remainder}
    R_\epsilon(Q,d) =  \sum_{n=2}^\infty {(i \epsilon)^n \over
    n!} \sum_{k=0}^n (-1)^{n-k} \left( \begin{array}{c} n \\ k \\
    \end{array} \right) \langle d \mid \hat{c}^{n-k} \hat{\rho}_s
    \hat{c}^k \mid d \rangle  \langle Q \mid \hat{P}^{n-k}
    \hat{\rho}_a \hat{P}^k \mid Q \rangle
\end{equation}
\end{widetext}
is a ``remainder term". The Lagrange form of the remainder term is
\begin{equation}
    R_\epsilon(Q,d) = {1 \over 2} {d^2 \rho_\xi (Q,d) \over d \xi^2}
    \epsilon^2
\end{equation}
for some $\xi$ in the range $(0,\epsilon)$. It can be shown that
an upper limit for the remainder term is given for $\xi=\epsilon$.
The lowest order approximation to the remainder term is found for
$\xi=0$

Assuming that the current density of the pointer vanishes
\begin{equation}\label{eq:zeroquantumcurrent}
    \langle Q \mid \hat{P} \hat{\rho}_a \mid Q \rangle +
    \langle Q \mid \hat{\rho}_a \hat{P} \mid Q \rangle = 0,
\end{equation}
it can be shown that
\begin{equation}
    \langle Q \mid \hat{\rho}_a \hat{P} \mid Q \rangle = {i \over
    2} {\partial \over \partial Q} \langle Q \mid \hat{\rho}_a
    \mid Q \rangle.
\end{equation}
We then find that
\begin{equation}\label{eq:jointdensity}
    \rho_\epsilon(Q,d) = \rho_o(Q,d) - \epsilon \mathrm{Re}
    \left ( c_w \right ) {\partial \rho_o(Q,d) \over \partial Q} +
    R_\epsilon(Q,d),
\end{equation}
where
\begin{equation}\label{eq:mixedweak}
    c_w(d) = {\langle d \mid \hat{c} \hat{\rho}_s \mid d
    \rangle \over \langle d \mid \hat{\rho}_s \mid d
    \rangle}
\end{equation}
can be recognized as the ``weak value" of the observable $\hat{c}$
for an object preselected in a mixed state $\hat{\rho}_s$ and
postselected in the eigenstate $\mid d \rangle$
\cite{Aharonov+AlbertETAL-ResuMeasCompSpin:88,%
Johansen-Whatvaluobsebetw:03},

By integrating Eq. (\ref{eq:jointdensity}), it is found that, to
first order in $\epsilon$, the probability density for the object
observable $d$ is unaffected by the measurement interaction,
\begin{equation}
    \int dQ \rho_\epsilon(Q,d) \approx \int dQ \rho_0 (Q,d) =
    \langle d \mid \hat{\rho}_s \mid d \rangle.
\end{equation}
We may write the conditional probability density for the pointer
position as
\begin{eqnarray}
    \rho_\epsilon(Q \mid d) = {\rho_\epsilon(Q,d) \over \int dQ
    \rho_\epsilon(Q,d)} \approx \hat{\cal T} \langle Q \mid
    \hat{\rho}_a \mid Q \rangle,
\end{eqnarray}
where
\begin{equation}
    \hat{\cal T} = 1 - \epsilon \mathrm{Re} (c_w) {\partial \over
    \partial Q}
\end{equation}
is a first order translation operator. The pointer position, given
a value $d$ of the object observable, has been translated by a
distance $\epsilon \mathrm{Re} (c_w) $.

If the standard deviation of the pointer position is $\sigma$, the
basic condition for a weak measurement is that the translation of
the pointer should be small compared to the standard deviation of
the pointer, $\mid \epsilon \mathrm{Re} \left ( c_w \right ) \mid
\ll \sigma$. For a comparison with the requirements for standard,
projective measurements, see Ref.
\cite{Haake+Walls-Overamplmetequan:87}. To be precise, also the
second and higher order corrections to the expectation value of
the pointer position should be small compared to the first order
change. This requires that
\begin{equation}
    \left | \int dQ \; Q \; R_\epsilon (Q,d) \right | \;
    \ll \; \left | \epsilon \mathrm{Re} \left ( c_w \right )
    \right | \langle d \mid \hat{\rho}_s \mid d \rangle.
\end{equation}

We now turn to weak measurements of $c$-number observables. This
is relevant, e.g., in classical mechanics and for classical
radiation fields. We assume that the object and pointer both can
be described by a classical phase space distribution. Prior to the
measurement interaction, we assume that the object plus pointer is
in a product state $F_0 = F_s(q,p) F_a(Q,P)$, where capital
letters denote the pointer. We consider a weak measurement of a
general $c$-number object variable $c(q,p)$, and assume that the
interaction Hamiltonian is
\begin{equation}\label{eq:classicalinteraction}
    H_\epsilon = \epsilon \, \delta(t) \, c(q,p) \, P.
\end{equation}
This is the $c$-number equivalent of the quantum interaction term
(\ref{eq:quantuminteraction}). We assume that the interaction
Hamiltonian dominates over any other terms in the Hamiltonian
during the short time of interaction. The equation of motion is
given by the classical Liouville's theorem
\begin{equation}\label{eq:liouville}
    {\partial F \over \partial t} = - \left \{ F, H_\epsilon \right\},
\end{equation}
where
\begin{equation}
    \left \{ F, H_\epsilon \right \} = \sum_i \left ( {\partial F
    \over \partial q_i} {\partial H_\epsilon \over \partial p_i} -
    {\partial F \over \partial p_i} {\partial H_\epsilon \over
    \partial q_i} \right ).
\end{equation}
For the Hamiltonian (\ref{eq:classicalinteraction}) we have
\begin{eqnarray}
    {\partial H_\epsilon \over \partial p} &=& \epsilon \, \delta(t) \,
    {\partial c(q,p) \over \partial p} \, P,\\
    {\partial H_\epsilon \over \partial q} &=& \epsilon \,\ \delta(t) \,
    {\partial c(q,p) \over \partial q} \, P,\\
    {\partial H_\epsilon \over \partial P} &=& \epsilon \, \delta(t) \,
    c(q,p),\\
    {\partial H_\epsilon \over \partial Q} &=& 0.
\end{eqnarray}
The Poisson bracket therefore may be written as
\begin{equation}
    \left \{ H_\epsilon, F \right \} = \epsilon \, \delta(t) \,
    \left [ P \, \left ( {\partial c \over \partial p} {\partial
    \over \partial q} - {\partial c \over \partial q}
    {\partial  \over \partial p} \right ) + c {\partial  \over
    \partial Q} \right ] F.
\end{equation}
This has the form $\left \{ F, H_\epsilon \right \} = \hat{\cal
H}_\epsilon F$. Liouville's theorem (\ref{eq:liouville}) then can
be written in a form similar to the Schr{\"o}dinger equation,
\begin{equation}
    {\partial F \over \partial t} = - \hat{\cal H}_\epsilon F.
\end{equation}
The state after the measurement interaction can then be expressed
as $F_\epsilon = \hat{\cal U}_\epsilon F_0$, where the classical
propagator is
\begin{equation}\label{eq:classicalpropagator}
\hat{\cal U}_\epsilon = e^{- \int \hat{\cal H}_\epsilon dt}.
\end{equation}
In the particular problem considered here, we write $\hat{\cal
H}_\epsilon = \epsilon \delta (t) \hat{{\cal K}}$, where
\begin{equation}
    \hat{{\cal K}} =  P \left (
    {\partial c \over \partial p} {\partial
    \over \partial q} - {\partial c \over \partial q} {\partial
    \over \partial p} \right ) + c {\partial \over \partial Q}.
\end{equation}
The propagator (\ref{eq:classicalpropagator}) then simplifies to
\begin{equation}
    \hat{\cal U}_\epsilon = e^{- \epsilon \hat{{\cal K}}}.
\end{equation}
We consider this propagator to first order in $\epsilon$,
$\hat{\cal U}_\epsilon \approx 1 - \epsilon \hat{{\cal K}}$. After
the measurement interaction, the joint probability density for $Q$
and $q$ is
\begin{equation}
    \rho_\epsilon(Q, q) = \int dp \int d P \; F_\epsilon(q,p,Q,P).
\end{equation}
We then find that
\begin{widetext}
\begin{eqnarray}
    \rho_\epsilon(Q,q) &\approx&  \rho_0(Q,q) + \epsilon \int dp
    \left [ {\partial c(q,p) \over \partial q}
    {\partial F_s(q,p) \over \partial p} - {\partial c(q,p) \over
    \partial p} {\partial F_s(q,p) \over \partial q} \right ]
    \int dP P F_a(Q,P) \nonumber \\
    &-& \epsilon \int dp \; c(q,p) F_s(q,p)\;  {\partial f_a(Q)
    \over \partial Q}.
\end{eqnarray}
\end{widetext}
where
\begin{eqnarray}
    \rho_0(Q,q) &=& f_s(q) f_a(Q),\\
    f_s(q) &=& \int dp F_s(q,p), \\
    f_a(Q) &=& \int dp F_a(Q,P).
\end{eqnarray}
We again assume that the current density of the pointer vanishes,
\begin{equation}
    \int dP \; P \; F_a(Q,P) = 0.
\end{equation}
It then follows that to the first order in $\epsilon$,
\begin{equation}\label{eq:classicaljoint}
    \rho_\epsilon(Q, q) = \hat{\cal T}_c \rho_0(Q,q),
\end{equation}
where
\begin{equation}
    \hat{{\cal T}_c} = 1 - \epsilon c_w
    {\partial \over \partial Q}
\end{equation}
is once more a translation operator, and where
\begin{equation}\label{eq:classicalweakvalue}
    c_w(q) = {\int dp \; c(q,p) \; F_s(q,p),\\
    \over \int dp \, F_s(q,p)}
\end{equation}
is the weak value of the $c$-number observable $c(q,p)$. We see
that this is the conditional expectation value of $c(q,p)$. In
other words, $c_w$ is simply the expectation value of $c(q,p)$
``given" $q$.

By integrating Eq. (\ref{eq:classicaljoint}) over $Q$, it follows
that
\begin{equation}
    f_s(q) = \int dQ \rho_\epsilon(Q, q).
\end{equation}
This shows that to the first order in $\epsilon$, the probability
density of $q$ is unaffected by the measurement interaction. The
conditional probability density for the pointer position then
reads
\begin{equation}
    \rho_\epsilon(Q \mid q) = {\rho_\epsilon(Q, q) \over \int
    dQ \rho_\epsilon(Q, q)} = \hat{{\cal T}_c} f_a(Q).
\end{equation}
This shows that the pointer position $Q$ has been translated by a
distance $\epsilon c_w$. Also in this case, the measurement can be
considered to be weak if $\mid \epsilon c_w \mid \ll \sigma$,
where $\sigma$ is the initial position uncertainty of the pointer.

In conclusion, it was found that weak measurements can be
performed with a much wider class of pointer states than thought
previously. Any state of the pointer can be used provided that the
current density of the pointer vanishes. This conclusion is valid
regardless of whether the observables are noncommuting or whether
they are $c$-numbers.

\end{document}